\documentclass[prd,twocolumn,preprintnumbers,amsmath,amssymb,nofootinbib,english,showpacs]{revtex4-1}

\usepackage{mathrsfs}
\usepackage{graphics}
\usepackage{epsfig}
 \usepackage{graphicx}
\usepackage{bm}
\usepackage{hyperref}
\usepackage[usenames]{color}
\usepackage{url}

\hypersetup{
    colorlinks=true,
    linkcolor=red,
    citecolor=red,
}

\newcommand{\dr}{\textcolor{black}}
\newcommand{\tc}{\textcolor{black}}
\newcommand{\mv}{\textcolor{black}}

\newcommand{\dww}{\textcolor{black}}
\newcommand{\dw}{}

\newcommand{\remove}[1]{}

\def\etal{{\frenchspacing\it et al.}}
\def\ie{{\frenchspacing\it i.e.}}
\def\eg{{\frenchspacing\it e.g.}}

\def\be{\begin{equation}}
\def\ee{\end{equation}}
\def\ba{\begin{eqnarray}}
\def\ea{\end{eqnarray}}

\def\d{\rm d}

\def\p{{\rm prior}}
\def\fid{{\rm fid}}

\begin{document}

\title{Reconstruction of the dark matter-vacuum energy interaction}

\author{Yuting Wang$^{1,2}$, Gong-Bo Zhao$^{1,2}$, David Wands$^{2}$, Levon Pogosian$^{3}$, Robert G. Crittenden$^{2}$}

\affiliation{$^{1}$National Astronomy Observatories, Chinese Academy of Science, Beijing, 100012, People's Republic of China}

\affiliation{$^{2}$Institute of Cosmology and Gravitation, University of Portsmouth, Portsmouth, PO1 3FX, United Kingdom}

\affiliation{$^{3}$Department of Physics, Simon Fraser University, Burnaby, British Columbia, Canada V5A 1S6}

\begin{abstract}

An interaction between the vacuum energy and dark matter is an intriguing possibility which may offer a way of solving the cosmological constant problem. Adopting a general prescription for momentum exchange between the two dark components, we reconstruct $\alpha(a)$, the temporal evolution of the coupling strength between dark matter and vacuum energy, in a nonparametric Bayesian approach using combined observational data sets from the cosmic microwave background, supernovae and large scale structure. An evolving interaction between the vacuum energy and dark matter removes some of the tensions between different data sets. However, it is not preferred over $\Lambda$CDM in the Bayesian sense, as improvement in the fit is not sufficient to compensate for the increase in the volume of the parameter space.
\end{abstract}


\maketitle

\section{Introduction}

The discovery of cosmic acceleration \cite{Riess:1998May, Perlmuter:1999Dec} has inspired the development of a wide range of dark energy (DE) and modified gravity models. The simplest DE candidate, the cosmological constant $\Lambda$, is a parameter in General Relativity and is consistent with all current observations \cite{Planck:2015Cos}. The main problem with $\Lambda$ is not why it has a particular value, but the fact that the vacuum energy contribution to $\Lambda$ is sensitive to the ultraviolet cutoff scale and requires a technically unnatural fine-tuning \cite{Burgess:2013ara}, which is the long-standing cosmological constant problem (CCP) \cite{Weinberg:1988cp}. Most of the dynamical DE and modified gravity models proposed in the literature do not offer a solution to this old problem. The CCP would be surmountable if there was a dynamical mechanism by which vacuum energy could decay from its initially large value and settle at an attractor consistent with the observed value of $\Lambda$. While a full theory of the quantum vacuum that would contain such a mechanism does not exist, this idea has motivated phenomenological models of decaying vacuum energy \cite{Bertolami:1986bg,Freese:1986dd,Chen:1990jw,Carvalho:1991ut,Berman:1991zz,Shapiro:2000dz,Sola:2011qr}. 

One way to have a nonconstant vacuum energy is to introduce a new dynamical degree of freedom, \eg, a scalar field. An alternative approach, which avoids explicitly introducing new degrees of freedom while still preserving the general covariance of the evolution of cosmological perturbations, is to allow for the exchange of momentum between the vacuum energy and other species \cite{Wands:2012Mar}. Both the minimally \cite{Ratra:1998apj, Ratra:1998prd, Caldwell:1997ii} and nonminimally \cite{Wetterich:1994bg, Holden:1999hm, Amendola:1999er} coupled quintessence models can be cast in this general framework. In the former case, the vacuum (the potential energy) exchanges energy with the kinetic energy of the scalar field. In the latter, there is an additional exchange with matter. In this sense, one can say that a time-dependent vacuum energy is necessarily interacting. Since additional interactions in the visible matter sector are strongly constrained while the nature of dark matter (DM) is largely unknown, we will consider models in which the vacuum interacts only with DM.

In this paper we adopt the phenomenological model of vacuum energy evolution introduced in \cite{Wands:2012Mar} which avoids dealing with explicit additional degrees of freedom. The vacuum equation of state is by definition equal to  $-1$, but the vacuum energy density $V$ is allowed to vary because of the interaction with DM, $\nabla_{\mu}V=-Q_{\mu}$.

\section{The model} 

In the interacting vacuum energy model the background DM and vacuum energy densities obey continuity equations
\ba 
\label{rhodmdot} \dot{\rho}_{\rm \tc{dm}}+3H\rho_{\rm \tc{dm}} = -Q \,; \ \dot{V} = Q \,,
\label{Vdot} 
\ea
where $Q$ denotes the energy transfer between DM and vacuum energy. 
\dw{An arbitrary energy transfer $Q$ can in principle reproduce an arbitrary background cosmology, with energy density $\rho=\rho_{\rm \tc{dm}}+V$ and pressure $P=-V$ \cite{Wands:2012Mar}. This reduces to $\Lambda$CDM when $Q=0$.}

To calculate the linear perturbations, we need to specify a covariant form of the energy-momentum transfer 4-vector. Following \cite{Wang:2013Jan, Wang:2014Apr}, we assume that the covariant interaction is parallel to the 4-velocity of DM, $Q^{\mu} = {Q}u^{\mu}_{(\mathrm{dm})}$. There are other choices \cite{Wands:2012Mar}, but this covariant form for the interaction means that there is no momentum transfer in the DM \dww{comoving-orthogonal frame} and hence the DM particles follow geodesics, as in $\Lambda$CDM. Although the interacting vacuum does allow inhomogeneous perturbations, one can always choose a frame in which the vacuum is spatially homogeneous, $\delta V=0$. For geodesic DM this coincides with comoving-orthogonal frame \cite{Wang:2013Jan, Wang:2014Apr} and the perturbation equations are particularly simple in this gauge. Note that in this case the energy transfer is a potential flow, $Q^\mu = -\nabla^\mu V$, and therefore the matter velocity $u^{\mu}_{(\mathrm{dm})}$ must be irrotational \cite{Sawicki:2013wja}. The DM density contrast evolves according to 
\begin{eqnarray}
 \label{dmenergy}
 \dot{\delta}_{\rm \tc{dm}}=-\vartheta+\frac{Q(a_i)}{\rho_{\rm \tc{dm}}}\delta_{\rm \tc{dm}} \,,
\end{eqnarray}
where the divergence of the matter 3-velocity is given by the extrinsic curvature of the metric, $\vartheta=\dot{h}/2$ in the comoving-synchronous gauge, and $h$ is the scalar mode of metric perturbations \cite{Ma:1995Jun}. 

We take $Q$ to be of a form inspired by the generalized Chaplygin gas model \cite{Kamenshchik:2001cp,Bento:2002ps}, $Q = 3\alpha H \rho_{\rm \tc{dm}} V / (\rho_{\rm \tc{dm}} +V)$, where $\alpha$ is a dimensionless coupling parameter \cite{Wands:2012Mar}. This form naturally reproduces a conventional matter-dominated solution at early times, and a vacuum dominated solution at late times, while allowing more general evolution in between. Previous studies \cite{Wang:2013Jan, Wang:2014Apr} have constrained the interaction assuming $\alpha = {\rm const.}$\footnote{Salvatelli \etal \cite{Salvatelli:2014Jun} considered a related dimensionless parameter, $q=-3\alpha\rho_{\rm \tc{dm}}/(\rho_{\rm \tc{dm}} +V)$, in four redshift intervals.}

In this paper, we make no assumptions about the time dependence of $\alpha$ and directly reconstruct it from data using the nonparametric Bayesian approach introduced in \cite{Crittenden:2011Dec}. We are thus able to describe a general background cosmology and reproduce any equation of state seen in quintessence models, but the perturbations are of a restricted form with vanishing sound speed but nonzero energy transfer to or from DM, determined by the background cosmology. Thus our model is degenerate with quintessence models in terms of the background cosmology, but distinguished by the evolution of perturbations.

\section{The reconstruction method}

We first discretize $\alpha(a)$ into bins $\alpha_i = \alpha(a_i)$, $i=1,...,N$, distributed uniformly in the interval $[a_{\rm min}, a_{\rm max}]$, giving us $N$ parameters $\alpha_i$ that can be fit to data. Any representation of a function with a finite number of parameters is necessarily \dr{inaccurate} and, in the absence of additional priors, the outcome of the fit would directly depend on $N$. As shown by Crittenden \etal \,\cite{Crittenden:2011Dec}, it is possible to eliminate the dependence on $N$ and explicitly control the reconstruction bias by adding a prior that correlates the nearby bins.

In the correlated prior approach one assumes that $\alpha(a)$ is a smooth function, so that its values at neighboring points in $a$ are not entirely independent. More specifically, $\alpha(a)$ is taken to be a Gaussian random variable with a given correlation $\xi$ between its values at $a$ and $a'$, $\xi (|a - a'|) \equiv \left\langle [\alpha(a) - \alpha^{\rm fid}(a)][\alpha(a') - \alpha^{\rm fid}(a')] \right\rangle$, which is nonzero for $|a-a'|$ below a given ``correlation length'' but vanishes at much larger separations, and $\alpha^{\rm fid}(a)$ is a reference fiducial model. Given a particular functional form of $\xi (|a - a'|)$, one can calculate the corresponding covariance matrix for the $N$ parameters $\alpha_i$:
\be
\label{eq:Cij} C_{ij}=\frac{1}{\Delta^2}\int_{a_i}^{a_i+\Delta}{\d}a \int_{a_j}^{a_j+\Delta}{\d}a'~\xi (|a - a'|) \,,
\ee 
where $\Delta$ is the bin width. This covariance matrix defines a (Gaussian) prior probability distribution for parameters $\alpha_i$ and its product with the likelihood, according to Bayes' theorem, gives the desired posterior distribution. Thus, the reconstruction amounts to finding the minimum of $\chi^2=\chi_{\rm prior}^2+\chi_{\rm data}^2$, where $\chi_{\p}^2=\left(\mathbf{\alpha}-\mathbf{\alpha}^{\fid}\right)^T \mathbf{C}^{-1} \left(\mathbf{\alpha}-\mathbf{\alpha}^{\fid}\right)$ and $\mathbf{\alpha}^{\fid}$ is 
some fiducial model. To avoid dependence on $\mathbf{\alpha}^{\fid}$, we take it to be constant and marginalize over its value following \cite{Crittenden:2011Dec}. 

We adopt the CPZ form \cite{Crittenden:2011Dec, Crittenden:2009Dec} for the correlation function, $ \xi(|a - a'|)=\xi(0)/[1+(|a - a'|/a_c)^2] $, where $a_c$ determines the correlation length and $\xi(0)$ sets the strength of the prior. This form has the advantage of making it possible to evaluate integrals in Eq.~(\ref{eq:Cij}) analytically, while also having a transparent dependence on its two parameters. We also note that, as shown in \cite{Crittenden:2011Dec}, the outcome of reconstruction is largely insensitive to the particular function chosen to represent $\xi(|a - a'|)$.

The correlation length $a_c$ sets the effective number of degrees of freedom allowed by the prior: $N_{\rm eff} \simeq (a_{\rm max}-a_{\rm min})/a_c$. As long as $N>N_{\rm eff}$, the reconstructed result is independent of $N$. Rather than adjusting $\xi(0)$ to control the strength of the prior, we use the variance of the mean given by $\sigma^2_{\bar{\alpha}} \simeq \pi\xi(0){a_c}/(a_{\rm max}-a_{\rm min})$ in the limit $a_c\ll a_{\rm max}-a_{\rm min}$ \cite{Crittenden:2011Dec, Crittenden:2009Dec}.
Namely, given $a_c$, one can adjust $\xi(0)$ to keep the prior on the uncertainty in the mean of $\alpha(a)$ independent of $N_{\rm eff}$. In what follows, we take $\sigma_{\bar{\alpha}}=0.04$ based on the constraint on a constant $\alpha$ obtained in 
\cite{Wang:2014Apr}, and set $a_c=0.06$. As we show later using principal component analysis (PCA), this choice of $a_c$ effectively separates the signal from noise, thus allowing a high-resolution reconstruction with minimal contamination from the noise.   

\begin{figure*}[htp]
\begin{center}
\includegraphics[scale=0.39]{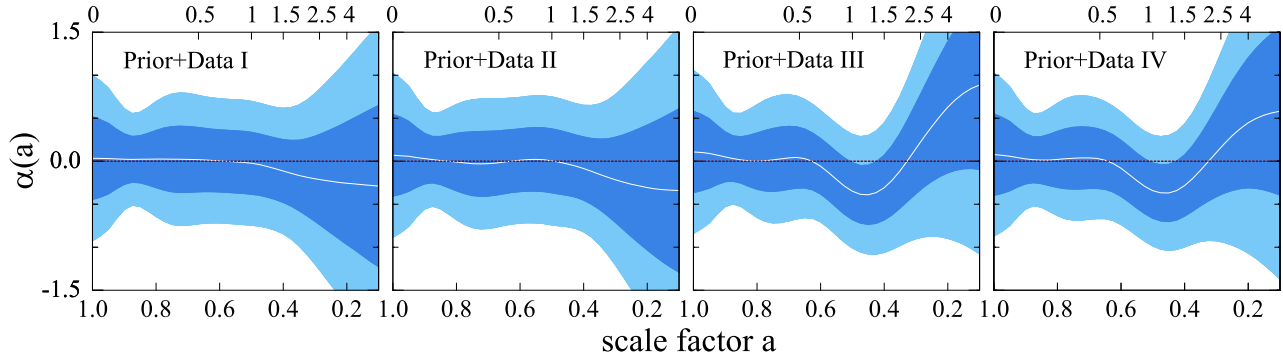}
\setlength{\abovecaptionskip}{-2.9cm}
\caption{Reconstruction of $\alpha(a)$ from four different data combinations: Data I: CMB+SN+BAO (without LyaF)+RSD (with AP effect); Data II: CMB+SN+BAO (without LyaF)+RSD (All); Data III: CMB+SN+BAO (All)+RSD (with AP effect) and Data IV: CMB+SN+BAO (All)+RSD (All). The \dr{best-fit} model (central white solid curves) with the 68, 95\% CL errors (dark and light blue shaded bands) are shown in each panel. \dr{We denote the redshift in each panel on the top x-axis.} The horizontal dashed line denotes the $\Lambda$CDM model. }\label{fig:alpha}
\end{center}
\end{figure*}

\section{Data}

Our data sets include the CMB temperature and polarization power spectra from Planck \cite{Planck:2013Mar:cos} and WMAP9 \cite{WMAP9:2012} respectively; the JLA supernovae sample \cite{Sako:2014Jan}; the BAO measurements of 6dFGRS \cite{Beutler:2011Jun}, SDSS DR7 \cite{Padmanabhan:2012Feb}, BOSS LOWZ \cite{Anderson:2013Dec} and BOSS Lyman-$\alpha$ Forest (LyaF) \cite{Aubourg:2014Nov}; the redshift space distortion (RSD) measurements which probe both the expansion and the growth history, namely ($D_V/r_s(z_d), F_{\rm AP}, f\sigma_8$) from BOSS CMASS \cite{Beutler:2013Dec} and ($A, F_{\rm AP}, f\sigma_8$) from WiggleZ \cite{Blake:2012Apr}, where $F_{\rm AP}$ quantifies the ``Alcock-Paczynski" effect \cite{AP}. We denote these data sets as ``RSD (with AP effect)," while we also use a ``RSD (All)" data set that includes additional RSD measurements without the AP effect: 6dFGRS \cite{Beutler:2012Apr}, 2dFGRS \cite{Percival:2004Jun}, SDSS LRG \cite{Samushia:2011Feb} and VIPERS \cite{Torre:2013Mar}. We present the constraints from four data combinations as shown in Fig.~\ref{fig:alpha}.

The RSD measurements constrain $(f\sigma_8)^2$, the product of the growth rate and the variance of the total matter density contrast. Note that the continuity equation of DM density contrast in our model  [Eq.~(\ref{dmenergy})] is different from that in $\Lambda$CDM and, therefore, the growth rate probed by the RSD is no longer simply that of the DM component. Namely, in $\Lambda$CDM, the velocity divergence of DM, $\vartheta\equiv\vec{\nabla}\cdot\vec{v}$, is simply $f_{\rm dm} H\delta_{\mathrm{dm}}$ where $f_{\mathrm{dm}} \equiv d \ln \delta_{\mathrm{dm}} /d \ln a$ is the DM growth rate. In the interacting vacuum model, Eq.~(\ref{dmenergy}) implies that 
\be
\nonumber
\vartheta = -\left[f_{\mathrm{dm}}-\frac{Q(a_i)}{H\rho_{\mathrm{dm}}}\right]H\delta_{\mathrm{dm}} \equiv -f_i H\delta_{\mathrm{dm}} \equiv -f_{\vartheta}H\delta_{\rm \tc{m}} \,.
\ee 
If $Q(a_i)\neq0$, the effective DM growth rate $f_i$, the appropriately weighted total matter growth rate $f_{\vartheta}$ \cite{BruniWandsinprep} and $f_{\rm dm}$ are different from each other. It can be shown that
\be 
\label{eq:fvartheta} f_{\vartheta}=\left( \frac{1}{f_i}\frac{\rho_{\rm \tc{dm}}}{\rho_{\rm \tc{m}}}+\frac{1}{f_{\rm \tc{b}}}\frac{\rho_{\rm \tc{b}}}{\rho_{\rm \tc{m}}}\right)^{-1}\,,
\ee
where $f_{\rm b}\equiv {d \ln \delta_{\rm b}}/{d \ln a}$ is the growth rate of baryons. In the absence of an interaction we recover the standard result: $f_{\vartheta}=f_i=f_{\rm \tc{b}}$, while, in general, it is $(f_{\vartheta}\sigma_8)^2$ that is measured by RSD.

\section{Results}

We discretize $\alpha(a)$ into $40$ bins uniform in $a$ within [0.001, 1], and use Monte Carlo Markov chains (MCMC) implemented in a modified version of {\tt CosmoMC} \cite{ref:MCMC} to fit them to data along with all other relevant cosmological parameters. A total $\chi^2$ is minimized and the joint posterior probability distribution for all the parameters is obtained after the MCMC has converged.  

\begin{figure*}[htp]
\begin{center}
\includegraphics[scale=0.45]{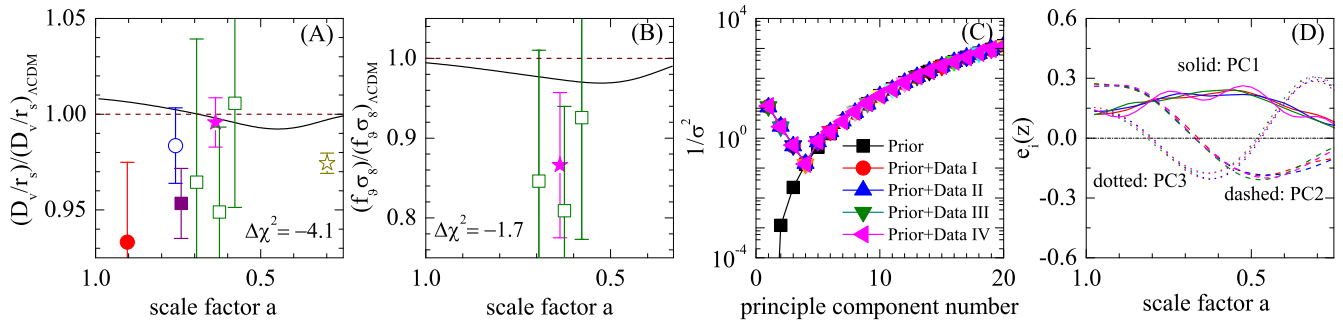}
\setlength{\abovecaptionskip}{-2.0cm}
\caption{Panels (A) and (B): The theoretical prediction of $D_V/r_s$ and $f_{\vartheta}\sigma_8$ by the best-fit $\alpha(a)$ model (solid line), and measurements (data with error bars), rescaled by the best-fit $\Lambda$CDM model. The rescaled BAO measurements are: 6dFGRS (filled circles), BOSS LOWZ (open circles), SDSS DR7 (filled squares), WiggleZ (open squares), CMASS (filled star), and LyaF (open star). The RSD points from WiggleZ (open squares) and CMASS (filled star) are shown in panel (B). The dashed horizontal line denotes the $\Lambda$CDM model. Panel (C): The eigenvalues of the covariance matrix obtained using data plus prior for four different data combinations, and using prior alone. Panel (D): The first three eigenvectors of the best-fit $\alpha(a)$. See the text for more details.}\label{fig:bao-fs8-pca}
\end{center}
\end{figure*}

The best-fit reconstructed models of $\alpha(a)$ (with 68\% and 95\% C.L. errors) are shown in Fig.~\ref{fig:alpha}. The $\Lambda$CDM fits the observations well when the LyaF BAO measurements are not included (Data I) and the reconstruction remains almost unchanged after adding more RSD data (Data II). However, when the LyaF data is included, we see evidence for an evolving $\alpha$: Data III (IV) shows a $1.8 (1.9\sigma)$ improvement in the fit for the $\alpha\ne0$ model. In these cases, the best-fit $\alpha(a)$ model changes sign during its evolution, \ie, it is positive at $z\gtrsim2.1$, implying an energy transfer from DM to vacuum energy, but negative at $0.6\lesssim z\lesssim2.1$, implying vacuum decay. At $z\lesssim0.6$, $\alpha$ is consistent with $\Lambda$CDM. The variations at higher and lower redshifts compensate to make the deceleration-acceleration transition redshift, $z_{\rm t}=0.6$ from Data III, close to the $\Lambda$CDM value of $z_{\rm t}=0.65$, which agrees with the value extracted from the $H(z)$ data \cite{Farooq:2013hq} using a Gaussian prior of $H_0=68\pm2.8\,{\rm km\,s^{-1}Mpc^{-1}}$, but is smaller than the value extracted using a prior of $H_0=73.8\pm2.4\,{\rm km\,s^{-1}Mpc^{-1}}$ \cite{Farooq:2013hq}. Again, adding more RSD data points (Data IV versus Data III) does not change the reconstruction significantly.

To understand this result,  in Fig.~\ref{fig:bao-fs8-pca} we show the theoretical predictions for $D_V/r_s$ and $f_{\vartheta}\sigma_8$ for the best-fit binned model from Data III rescaled by the corresponding best-fit $\Lambda$CDM model predictions, together with the actual measurements. In panel (A), the LyaF measurement (open star), has the smallest error bar and pulls the fit down, also making it more consistent with the CMASS measurement (filled star). This reduces the BAO $\chi^2$ by $4.1$. This fit is favored by the $f_{\vartheta}\sigma_8$ as well (panel B), further reducing the $\chi^2$ by $1.7$. The remaining data from CMB and SN slightly disfavor this model; namely, they increase $\chi^2$ by $\sim2$, but not enough to compensate for the reduction in $\chi^2$ from BAO and RSD. 

\begin{figure}[tbp]
\includegraphics[scale=0.32]{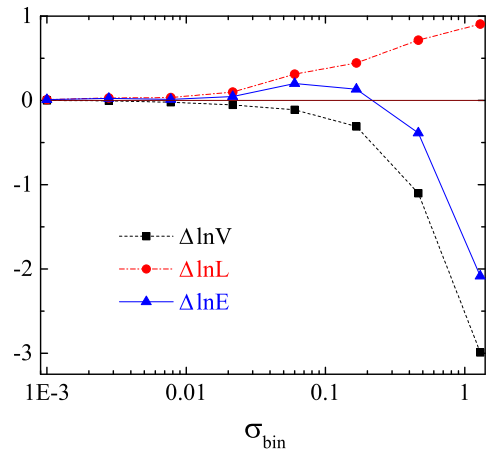}
\caption{The logarithms of the ratios of volumes of parameter space ($\Delta \ln \rm V$), of likelihoods ($\Delta \ln \rm L$), and of the evidences ($\Delta \ln \rm E$) for the binned model and the $\Lambda$CDM as a function of the strength of the prior set by the variance in each bin, $\sigma_{\rm bin}$.}\label{fig:bayes-evi}
\end{figure}

\mv{The improvement in the fit achieved by the binned model must be weighed against the increased number of degrees of freedom. This can be quantified via the Bayes factor, or the ratio of the Bayesian evidences of the interacting model and the $\Lambda$CDM model. For one model to be preferred over the other, the Bayes factor should be significantly greater than 1.
Since the evidence depends on the prior assumed for the binned coupling $\alpha(a_i)$, we follow the method in \cite{Zhao:2012Jul} to examine the dependence of the Bayes factor on the the variance in each bin, $\sigma_{\rm bin}$, which controls the strength of the prior.  In Fig.~\ref{fig:bayes-evi}, we plot the logarithm of the evidence ratio ($\Delta \ln \rm E$) as well as the logarithms of the ratios of the volumes of parameter space ($\Delta \ln \rm V$) and the likelihood ($\Delta \ln \rm L$) as a function of $\sigma_{\rm bin}$. For $\sigma_{\rm bin} \rightarrow 0$, the prior effectively forces $\alpha(a_i)$ to become equal to the fiducial model, which is $\Lambda$CDM, and all three ratios approach unity. Increasing $\sigma_{\rm bin}$ allows for more freedom in the variation of $\alpha(a_i)$ and improves the fit. However, as evident from Fig.~\ref{fig:bayes-evi}, the improvement in the fit is not sufficient to compensate for the increase in the parameter volume. The Bayes factor is only marginally greater than 1 for a limited range of $\sigma_{\rm bin}$ around the value of $0.1$, which was the value used in the reconstruction in Fig.~\ref{fig:alpha}. Thus, we conclude that the interacting model is not preferred over $\Lambda$CDM}.

\mv{Even with the lack of strong preference for the evolving model, it is still interesting to know to what extent our reconstruction constrains the evolution of the vacuum energy. For instance, one could ask which part of the information is informed by data and which part is informed by the prior. The errors shown in the reconstruction in Fig.~\ref{fig:alpha} are highly correlated, making their direct interpretation difficult. Principal component analysis (PCA) is a useful tool that can be used to analyze and tune the prior (for applications of PCA to DE studies, see \eg, \cite{Huterer:2002Jul,Crittenden:2009Dec,Crittenden:2011Dec}).} The PCA seeks the orthonormal eigenmodes of the inverse covariance matrix $F_{\alpha}$ of the $\alpha$ bins after marginalizing over other cosmological parameters. Namely, $F_{\alpha}=W^T\, \Lambda \,W$, with eigenvectors defined by the decomposition matrix $W$ and the eigenvalues given by the elements of the diagonal matrix $\Lambda$. Eigenmodes define independent linear combinations of the original parameters ($\alpha$ bins) that have uncorrelated errors and, thus, are easier to interpret. We can also use PCA to compare the eigenmodes with and without the prior which, as we explain below, allows us to estimate the effective number of degrees of freedom of $\alpha(a)$ constrained by data. Performing the PCA is straightforward, since $F_{\alpha}$ is one of the products of our MCMC calculation.

The eigenvalues of both the prior and the data+prior covariance are shown in panel (C) of Fig.~\ref{fig:bao-fs8-pca}. There are three data eigenmodes that are not affected by the prior, \ie, these three best constrained modes pass the prior with almost no penalty. The shapes of these modes are shown in panel (D)and we find that they are similar for all data combinations. The remaining modes, on the other hand, are dominated by the prior. Thus, even though our model has many bins of $\alpha$ as parameters, there are effectively only three additional degrees of freedom. \mv{As mentioned earlier, we find that the total $\chi^2$ is reduced by $\sim4$, which is somewhat greater than expected for a model with three additional parameters. }

The eigenvectors provide a natural basis onto which an arbitrary $\alpha(a)$ can be expanded, \ie, $\alpha(a)=\sum_i \beta_i e_i(a)$. Given any $\alpha(a)$, the coefficients $\beta$ can then be found using the orthogonality of the modes. The $\beta$'s corresponding to the three best constrained eigenmodes of the best-fit models from the four data combinations are listed in Table \ref{tab:coeff}. Since the uncertainty in the third eigenmode is large, adding more modes by relaxing the prior [either by reducing $\xi(0)$ or $a_c$] would not notably change the fit.

\begin{table}
\begin{center}
\begin{tabular}{c|c|c|c|c}
\hline \hline
                    &Data I & Data II & Data III & Data IV \\ \hline
                    &$-0.17\pm 0.30$ &
                      $-0.19 \pm 0.30$ &
                      $-0.34 \pm 0.30$&
                      $-0.40 \pm 0.29$ \\ 
$\beta_i$    &$0.34 \pm 0.61$&
                     $0.34 \pm 0.63$ &
                     $0.47 \pm 0.65$&
                     $0.48 \pm 0.64$ \\
                  &$-0.25 \pm 1.35$&
                    $-0.21 \pm 1.35$&
                    $-0.19 \pm 1.34$&
                    $-0.10 \pm 1.31$\\ 
\hline\hline
\end{tabular}
\caption{The coefficients (best-fit and 68\% C.L. uncertainty) of the first three best-determined modes, $\beta_i$, of the best-fit models using different data combinations.}\label{tab:coeff}
\end{center}
\end{table}

\section{Conclusion and discussions}

We performed a high-resolution reconstruction of $\alpha$, the coupling between DM and vacuum energy, as a function of the scale factor using the latest observations including CMB, SN, BAO and RSD. Our model is degenerate with standard quintessence models in terms of the background cosmology, but is distinguished by the growth of perturbations. We found that, when the BAO measurement using the BOSS LyaF sample \cite{Aubourg:2014Nov} is used, an evolving $\alpha$ is mildly favored by the joint data set. Interestingly, the best-fit $\alpha(a)$ model changes sign during its evolution: $\alpha>0$ at higher redshifts, implying an energy transfer from DM to vacuum energy, while $\alpha<0$ at lower redshifts, corresponding to a decaying vacuum energy. A PCA study of our result shows that we have extracted three informative eigenmodes from the data.

The LyaF BAO measurement, which is the best existing BAO measurement at such a high redshift, is in tension with the $\Lambda$CDM model at the $2-2.5\sigma$ level. As noted in \cite{Aubourg:2014Nov,Delubac:2014aqe}, it can be interpreted as favoring a DE component with a negative energy density at $z\sim2.3$. The RSD measurements from BOSS and WiggleZ are also in tension with $\Lambda$CDM: the RSD measurements favor a lower growth rate than the $\Lambda$CDM prediction at low redshifts. The interacting vacuum model provides another physical interpretation of these tensions if $\alpha$ is allowed to change sign during its evolution. We note that extracting BAO from LyaF data is a relatively new field, and the current measurement could be subject to systematic issues \cite{Aubourg:2014Nov}. Our method opens a new window into investigation of the interacting vacuum model that can be applied to improved future datasets as they become available. 

~
\acknowledgements{We thank M. Bruni, H. Borges and V. Salvatelli for discussions. Y. W. is supported by the NSFC grant No. 11403034 and the China Postdoctoral Science Foundation Grant No. 2014M550091. G. B. Z. is supported by the 1000 Young Talents program in China, by the Strategic Priority Research Program ``The Emergence of Cosmological Structures" of the CAS, Grant No. XDB09000000. L. P. is supported by NSERC and acknowledges the hospitality at the ICG. The work was supported by STFC Grants No. ST/H002774/1 and No. ST/L005573/1.}


\begin{thebibliography}{9}
\bibitem{Riess:1998May}
  A.~G.~Riess, {\it et al.},
  Astron.\ J.\ {\bf 116}, 1009 (1998).

\bibitem{Perlmuter:1999Dec}
  S.~Perlmutter, {\it et al.},
  Astrophys.\ J.\ {\bf 517}, 565 (1999).

\bibitem{Planck:2015Cos}
   P.~A.~R.~Ade, {\it et al.},
   arXiv:1502.01589.

\bibitem{Burgess:2013ara} 
  C.~P.~Burgess,
  arXiv:1309.4133 [hep-th].

\bibitem{Weinberg:1988cp} 
  S.~Weinberg,
  Rev.\ Mod.\ Phys.\  {\bf 61}, 1 (1989).

\bibitem{Bertolami:1986bg} 
  O.~Bertolami,
  Nuovo Cim.\ B {\bf 93}, 36 (1986).

\bibitem{Freese:1986dd} 
  K.~Freese, F.~C.~Adams, J.~A.~Frieman and E.~Mottola,
  Nucl.\ Phys.\ B {\bf 287}, 797 (1987).

\bibitem{Chen:1990jw} 
  W.~Chen and Y.~S.~Wu,
  Phys.\ Rev.\ D {\bf 41}, 695 (1990); [Erratum-ibid.\ D {\bf 45}, 4728 (1992)].

\bibitem{Carvalho:1991ut} 
  J.~C.~Carvalho, J.~A.~S.~Lima and I.~Waga,
  Phys.\ Rev.\ D {\bf 46}, 2404 (1992).

\bibitem{Berman:1991zz} 
  M.~S.~Berman,
  Phys.\ Rev.\ D {\bf 43}, 1075 (1991).

\bibitem{Shapiro:2000dz} 
  I.~L.~Shapiro and J.~Sola,
  JHEP {\bf 0202}, 006 (2002).

\bibitem{Sola:2011qr} 
  J.~Sola,
  J.\ Phys.\ Conf.\ Ser.\  {\bf 283}, 012033 (2011).
      
\bibitem{Wands:2012Mar}
  D.~Wands, J.~De-Santiago and Y.~Wang,
  Class.\ Quant.\ Grav.\  {\bf 29}, 145017 (2012).
 
\bibitem{Ratra:1998apj}
 B.~Ratra and P.~J.~E.~Peebles,
  Astrophys. J. 325, L17 (1988).
 
 
\bibitem{Ratra:1998prd}
  B.~Ratra and P.~J.~E.~Peebles, 
  Phys. Rev. D 37, 3406 (1988).


\bibitem{Caldwell:1997ii} 
  R.~R.~Caldwell, R.~Dave and P.~J.~Steinhardt,
  Phys.\ Rev.\ Lett.\  {\bf 80}, 1582 (1998).

\bibitem{Wetterich:1994bg} 
  C.~Wetterich,
  Astron.\ Astrophys.\  {\bf 301}, 321 (1995).

\bibitem{Holden:1999hm} 
  D.~J.~Holden and D.~Wands,
  Phys.\ Rev.\ D {\bf 61}, 043506 (2000).

\bibitem{Amendola:1999er} 
  L.~Amendola,
  Phys.\ Rev.\ D {\bf 62}, 043511 (2000).
 
\bibitem{Wang:2013Jan}
  Y.~Wang, D.~Wands, L.~Xu, J.~De-Santiago and A.~Hojjati,
  Phys.\ Rev.\ D {\bf 87}, 083503 (2013).

\bibitem{Wang:2014Apr}
  Y.~Wang, D.~Wands, G.~-B.~Zhao and L.~Xu,
  Phys.\ Rev.\ D {\bf 90}, 023502 (2014).
  
\bibitem{Sawicki:2013wja} 
  I.~Sawicki, V.~Marra and W.~Valkenburg,
  Phys.\ Rev.\ D {\bf 88}, 083520 (2013).
  
\bibitem{Ma:1995Jun}
  C.~P.~Ma and E.~Bertschinger,
  Astrophys.\ J.\ {\bf 455}, 7 (1995).
   
\bibitem{Kamenshchik:2001cp} 
  A.~Y.~Kamenshchik, U.~Moschella and V.~Pasquier,
  Phys.\ Lett.\ B {\bf 511}, 265 (2001).

\bibitem{Bento:2002ps} 
  M.~C.~Bento, O.~Bertolami and A.~A.~Sen,
  Phys.\ Rev.\ D {\bf 66}, 043507 (2002).
  
\bibitem{Salvatelli:2014Jun}
  V.~Salvatelli, N.~Said, M~.Bruni, A.~Melchiorri and D.~Wands,
  Phys.\ Rev.\ Lett.\ {\bf 113}, 181301 (2014).
  
\bibitem{Crittenden:2011Dec}
  R.~G.~Crittenden, G.~B.~Zhao, L.~Pogosian, L.~Samushia and X.~Zhang,
  JCAP {\bf 1202}, 048  (2012).
 
 
\bibitem{Crittenden:2009Dec}
  R.~G.~Crittenden, L.~Pogosian and G.~B.~Zhao,
  JCAP {\bf 0912} (2009) 025.
  
\bibitem{Planck:2013Mar:cos}
  P.~A.~R.~Ade {\it et al.}  [Planck Collaboration],
  Astron.\ Astrophys.\  {\bf 571}, A16 (2014).

\bibitem{WMAP9:2012}
 G.~Hinshaw {\it et al.},  
  Astrophys.\ J.\ Suppl.\  {\bf 208}, 19 (2013).

\bibitem{Sako:2014Jan}
  M.~Sako {\it et al.},  
  arXiv:1401.3317 [astro-ph.CO].
   
\bibitem{Beutler:2011Jun}
  F.~Beutler, {\it et al.},
  Mon.\ Not.\ Roy.\ Astron.\ Soc.\ {\bf 416}, 3017 (2011).

\bibitem{Padmanabhan:2012Feb}
  N.~Padmanabhan, {\it et al.},
  Mon.\ Not.\ Roy.\ Astron.\ Soc.\ {\bf 427}, 2132 (2012).

\bibitem{Anderson:2013Dec}
 L.~Anderson {\it et al.},  
  Mon.\ Not.\ Roy.\ Astron.\ Soc.\  {\bf 441}, 24 (2014).

\bibitem{Aubourg:2014Nov}
   $\acute{\mathrm{E}}$.~Aubourg, {\it et al.},
   arXiv:1411.1074 [astro-ph.CO].
 
\bibitem{Beutler:2013Dec}
  F.~Beutler {\it et al.}, 
  Mon.\ Not.\ Roy.\ Astron.\ Soc.\  {\bf 443}, no. 2, 1065 (2014).
    
\bibitem{Blake:2012Apr}
   C.~Blake, {\it et al.},
  Mon.\ Not.\ Roy.\ Astron.\ Soc.\  {\bf 425}, 405 (2012).

\bibitem{AP} 
  C.~Alcock and B.~Paczynski,
  Nature {\bf 281}, 358 (1979).
   
\bibitem{Beutler:2012Apr}
  F.~Beutler, {\it et al.},
  Mon.\ Not.\ Roy.\ Astron.\ Soc.\  {\bf 423}, 3430 (2012).
   
\bibitem{Percival:2004Jun}
   W.~J.~Percival, {\it et al.},
   Mon.\ Not.\ Roy.\ Astron.\ Soc.\ {\bf 353}, 1201 (2004).
   
\bibitem{Samushia:2011Feb}
   L.~Samushia, W.~J.~Percival and A.~Raccanelli,
   Mon.\ Not.\ Roy.\ Astron.\ Soc.\ {\bf 420}, 2102 (2012).
   
\bibitem{Torre:2013Mar}
  S.~de la Torre, {\it et al.},
  Astron.\ Astrophys.\  {\bf 557}, A54 (2013).
  
 \bibitem{BruniWandsinprep}
M.~Bruni, V.~Salvatelli and D.~Wands, {\em in preparation}.
   
\bibitem{ref:MCMC}
  A.~Lewis and S.~Bridle,
  Phys.\ Rev.\ D {\bf 66}, 103511 (2002).

\bibitem{Farooq:2013hq} 
  O.~Farooq and B.~Ratra,
  Astrophys.\ J.\  {\bf 766}, L7 (2013).
  
\bibitem{Huterer:2002Jul}
  D.~Huterer and G.~Starkman
  Phys.\ Rev.\ Lett.\ {\bf 90}, 031301 (2003).

\bibitem{Zhao:2012Jul}
 G.~B.~Zhao, R.~G.~Crittenden, L.~Pogosian and X.~Zhang,
  Phys.\ Rev.\ Lett {\bf 109}, 171301 (2012).

\bibitem{Delubac:2014aqe} 
  T.~Delubac {\it et al.}, 
  Astron.\ Astrophys.\  {\bf 574}, A59 (2015).

\end{thebibliography}
\end{document}